\documentclass[12pt]{article}
\usepackage{amssymb}
\usepackage{amsfonts}
\usepackage{amsmath}
\usepackage{graphicx}
\usepackage{color}
\usepackage{hyperref}

\def\Z{\mathbb Z}

\newcommand{\beq}{\begin{equation}}
\newcommand{\eeq}{\end{equation}}
\newcommand{\bea}{\begin{eqnarray}}
\newcommand{\eea}{\end{eqnarray}}

     \advance \textheight by \topskip
     \makeatletter
     \@addtoreset{equation}{section}

     \makeatother
\def\beq{\begin{equation}}
\def\eq{\end{equation}}

\def\12{\frac{1}{2}}

\begin{document}

\title{ {\bf Spectral singularities in ${\cal PT}$-symmetric
periodic finite-gap systems}}

\author{
{\bf Francisco Correa${}^{a}$ and Mikhail S. Plyushchay${}^{b}$}
\\
[4pt]
{\small \textit{${}^{a}$ Centro de Estudios
Cient\'{\i}ficos (CECs), Casilla 1469, Valdivia, Chile}}\\
{\small \textit{${}^{b}$ Departamento de F\'{\i}sica, Universidad de
Santiago de Chile, Casilla 307, Santiago 2,
Chile  }}\\
\sl{\small{E-mails: correa@cecs.cl, mikhail.plyushchay@usach.cl} }}
\date{}

\maketitle

\begin{abstract}

The origin of spectral singularities in finite-gap singly periodic
${\cal PT}$-symmetric quantum systems is investigated.  We show that
they emerge from a limit of  band-edge states in a doubly periodic
finite gap system when the imaginary period tends to infinity. In
this limit, the energy gaps are contracted and disappear, every pair
of band states of the same periodicity at the edges of a gap
coalesces and transforms into a singlet state in the continuum. As a
result, these spectral singularities turn out to be analogous to
those in the non-periodic systems, where they appear as zero-width
resonances. Under the change of
topology from a non-compact into a compact one, spectral
singularities in the class of periodic systems we study are
transformed into exceptional points. The specific degeneration
related to the presence of finite number of spectral singularities
and exceptional points is shown to be coherently reflected by a
hidden, bosonized nonlinear supersymmetry.
\end{abstract}

\section{Introduction}

The discovery of complex Hamiltonians with the combined space
reflection and time reversal (${\cal PT}$) symmetry, which have a
real spectrum \cite{B&B}, opened a new branch of quantum mechanics
\cite{reviews}. Recently, systems with ${\cal PT}$-symmetry have
gained a lot of attention motivated by a possibility of its
experimental observation in nature, particularly,  in optical
systems \cite{optics}. The ideas of ${\cal PT}$-symmetry have also
been applied to different areas, including
quantum field theory \cite{QFT}, gravitation \cite{grav}
and relativistic quantum mechanics
\cite{rqm}, among others.

An interesting peculiarity of non-Hermitian Hamiltonians is related
with the presence of exceptional points \cite{kato} and spectral
singularities \cite{Naimark} in their spectra. Exceptional points
are particular states in the \emph{discrete} spectrum of an operator
where two eigenvectors of different energies coalesce to form a
unique state. These states were studied in several contexts,
see, e.g.,
\cite{EP1,EP2,EP3,EP4,EP5,EP6,bqr} and references therein. Spectral
singularities have a nature similar to that of exceptional points,
but within the
\emph{continuous} spectrum of a non-Hermitian operator
\cite{Samsonov1, AMHMD, Longhi1}. The energy values of the spectral
singularities appear as poles in the resolvent of an operator, as
zeros of a Wronskian of the Jost solutions, as well as divergences
in the scattering matrix. In 2009, Mostafazadeh showed that spectral
singularities in \emph{non-periodic} complex potentials appear as
zero-width resonances \cite{AM1}. Although the implications and
applications of spectral singularities have been analyzed in several
directions
\cite{Samsonov2,Samsonov3,Samsonov4,Samsonov5,Andrianov:2010an,
AM2,AM3,AM4,AM5,Heiss:2011ad}, their meaning in complex
\emph{periodic} potentials remains, however, unknown.

In this article we study a certain class of ${\cal PT}$-symmetric
complex potentials in which a finite number of spectral
singularities does appear. Their origin is explained by analyzing a
related, more general family of doubly periodic quantum models which
belong to a class of finite-gap systems \cite{Finitegap}. When the
imaginary period (that can be treated as a hidden imaginary
parameter of the potential) tends to infinity, energy gaps shrink
and disappear, while the pairs of singlet band states of the same
periodicity at the edges of each gap coalesce and produce singlet
states inside the doubly degenerate continuum. These turn out to be
the spectral singularities. This peculiar phenomenon is shown to be
characteristic for complex potentials, there exists no analog for
finite-gap real potentials. We show that the appearance of finite
number of spectral singularities in the indicated class of
non-Hermitian systems can be associated with a presence of a hidden,
bosonized non-linear supersymmetry \cite{bosonized}. A
compactification of the system, by imposing  the appropriate
periodicity condition for the wave functions, discretizes the
spectrum and transforms spectral singularities into exceptional
points. This provides a unified explanation for spectral
singularities and exceptional points for certain class of related
$\mathcal{PT}$-symmetric systems.

The paper is organized as follows. In section \ref{sec1} we
construct a family of periodic complex potentials with spectral
singularities by applying Crum-Darboux transformations to a free
particle. Section \ref{sec2} is devoted to the description of
spectral singularities and
exceptional points in the spectra of the obtained related systems
with non-compact and compact topologies, respectively. The origin
of the same spectral singularities from a specific limit of
finite-gap systems is explained in section \ref{sec3}. In section
\ref{sec4} we show that a hidden supersymmetry is associated with
the presence of finite number of spectral singularities and
exceptional points. Discussion is presented in section \ref{sec5} .

\section{Free particle and complex Darboux transformations}\label{sec1}

Let us consider a free particle on the real line~\footnote{We work
in the units $\hbar=2m=1$.} $-\infty < x< \infty$,
\begin{equation}\label{free}
    H=-\frac{d^2}{dx^2} \, .
\end{equation}
The spectrum of the system is
continuous and is described by
the states
\begin{equation}\label{plane}
\psi^{\pm}(x)=e^{\pm i k x}, \quad k\geq 0 \,.
\end{equation}
For $k>0$, (\ref{plane}) are plane waves for doubly
degenerate energy levels with  $E_k=k^2>0$. A singlet state $\psi=1$
corresponds to $E_0=0$ at the bottom of the spectrum. From the
system (\ref{free}) one can construct complex periodic Hamiltonians
with finite number of spectral singularities by employing
Crum-Darboux transformations. The procedure is analogous to the
construction of reflectionless potentials with a finite number of
bound states \cite{MatSal}. To define such a transformation, we
introduce a complex operator
\begin{equation}\label{dop}
    {\cal D}_{\alpha,\beta}=\frac{d}{dx}+\alpha \tan
    \left(x+i\delta \right)-\beta
    \cot \left(x+i\delta \right) \, ,
\end{equation}
where $\delta$ is a real parameter, $0<\delta<\pi/2$, and  construct
a higher order differential operator
\begin{equation}\label{intfree}
    {\cal F}_{r,s}  =\prod_{j=1}^r{\cal D}_{j,u} ,
    \qquad
    u=u(r,s;j) = \,
\begin{cases} j-r+s  &  \text { if} \,\,\, j-r+s>0 \,,
\\
0 &  \text { if} \,\,\,\,  j-r+s \leq 0\,.
\end{cases}
\end{equation}
 The upper index of the ordered product corresponds here to the
first term on the left side while the lower index denotes the last
term on the right side of the product. The parameters $r$ and $s$
take here integer values, and without any loss of generality  we can
assume that $r>s$ (see next Section). Operator (\ref{intfree})
intertwines the free particle Hamiltonian $H$ with those of
nontrivial systems described by Hamiltonians
\begin{equation}\label{hrs}
    H_{r,s}=-\frac{d^2}{dx^2}+\frac{r(r+1)}{\cos^2
    (x+i\delta)}+\frac{s(s+1)}
    {\sin^2(x+i\delta)} \, ,
\end{equation}
\begin{equation}\label{intertw}
    {\cal F}_{r,s} H=H_{r,s} {\cal F}_{r,s}\,,
\end{equation}
where the free particle system
$H$ corresponds to the zero values
of the parameters, $H_{0,0}=H$.

\begin{figure}[h!]
\centering
\includegraphics[scale=1]
{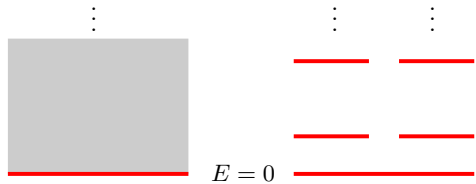}\caption{Spectrum of the free particle in the real
line (left) and in the compactified case (right). }\label{fig1}
\end{figure}

The nature of the continuous spectrum of $H_{0,0}$  can be modified
by changing the topology of the quantum problem. This is achieved by
compactifying the coordinate, $-\infty<x<+\infty\rightarrow 0 \leq
x<2\pi$, via the introduction of the
periodicity condition,
\begin{equation}\label{condi}
\psi^{\pm} (x+2\pi)=\psi^{\pm}(x) \,.
\end{equation}
This condition  is satisfied provided $k \in \mathbb{Z}$ in
(\ref{plane}), that transforms the energy spectrum of (\ref{free})
into the discrete one,
\begin{equation}
E_\ell=\ell^2,  \quad \ell=0,\pm1,\pm2,\pm3,\ldots\,.
\end{equation}
In this case we have an infinite set of discrete doubly degenerate
positive energy levels, while the ground state with $\ell=0$
($E_0=0$) is non-degenerate. Fig. \ref{fig1} illustrates the
spectrum in both cases, non-compact and compact ones.

In the next section we discuss general properties of the Hamiltonian
(\ref{hrs}) and the description of spectral singularities, where the
compactification scheme will be useful to understand  the relation
with exceptional points in the case of periodic systems. It is worth
to note that in the compactified case, the length of the space,
$2\pi$, is twice the period of (\ref{hrs}). This means that the
compactification condition (\ref{condi}) comprises (unifies) both
the periodic and anti-periodic Sturm-Liouville problems,
$\psi(0)=\pm \psi(\pi)$, $\psi'(0)=\pm \psi'(\pi)$, for the periodic
non-Hermitian system (\ref{hrs}). The same job is made, however,
just by imposing the periodicity condition $\psi(0)=\psi(2\pi)$.

\section{Spectral singularities of periodic ${\cal PT}$-symmetric potentials}
 \label{sec2}

Periodic potentials of complex nature
\cite{susypt1,BDM,jones,cervero,shin,samsonovroy,Sirre,kharemandal} are
of interest, particularly, in the context of optics and matter waves
physics, see for example refs. in \cite{Longhi2}. Hamiltonian
(\ref{hrs}) provides a good example of such a class of potentials,
for which the properties of non-Hermitian systems can be analyzed
using the advantage of exactly solvable systems.

The potential in (\ref{hrs}) has a real period $T=\pi$, and can be
treated as the complex version of the generalized trigonometric
P\"oschl-Teller potential. Although the potential is just a complex
shift in the coordinate of the Hermitian case, its spectral
properties are essentially different. Indeed, when $\delta=0$,  the
potential in (\ref{hrs}) is no longer complex, and has singularities
located at $x=l \pi$ and $x=(l+\frac{1}{2})\pi$, where $l \in
\mathbb{Z}$. Consequently, the quantum interpretation in the real
case is quite different in comparison with the complex one. The
presence of singularities implies that the particle is confined in
the region between the adjacent singularities, without tunneling
through them. Therefore, one ought to impose the boundary conditions
by requiring that the wave functions vanish at singular points. This
results in an infinite number of bound states which correspond to
non-degenerate energy levels. The picture is the same as in the case
of the infinite square well potential. Moreover, both systems are
related by the Crum-Darboux transformations (\ref{intfree}) when
$\delta=0$, remembering that the infinite square well potential
problem is no more than the free particle system subjected to the
specific boundary conditions. In the Hermitian case, the quantum
interpretation of a particle with the compactified coordinate
requires also the necessary boundary conditions of vanishing of the
wave functions at the singular points of the potentials.

For $\delta \neq 0$ the situation radically changes. The
singularities on the real line disappear, and the simultaneous
action of the parity ${\cal P}$, ${\cal P}x{\cal P}=-x$, and time
reversal ${\cal T}$, ${\cal T}i{\cal T}=-i$, operators leaves the
Hamiltonian invariant. Before explaining the spectral properties of
the family of systems  (\ref{hrs}), we review some general aspects
of the Hamiltonian $H_{r,s}$.

The shift $x \rightarrow x+\pi/2$ in the Hamiltonian (\ref{hrs})
interchanges the parameters $r$ and $s$, $H_{r,s}
(x+\pi/2)=H_{s,r}(x)$, without affecting the spectral
characteristics of the system. Since the systems $H_{r,s}$ and
$H_{s,r}$ can also be related by Crum-Darboux transformations
constructed in terms of operators of the form (\ref{dop}), we say
that they are self-isospectral \cite{self}.

This fact, in addition to the symmetries $r \rightarrow -r-1$ and $s
\rightarrow - s-1$, tells us that we can consider just non-negative
integer values of the parameters. The case when $r=s=l$ is a
particular case which is presented equivalently as
\begin{equation}
    H_{l,l}(x+\pi/4)=4H_{l,0}(\chi)=4\left(-\frac{d^2}
    {d\chi^2}+\frac{l(l+1)} {\cos^2(\chi+2i\delta)}
    \right)\,,
\end{equation}
where $\chi=2x$. Hence, the spectrum of the Hamiltonian (\ref{hrs})
with $r=s$ is given by rescaling of that for the case with $s=0$,
and then it is indeed sufficient to
consider the cases $r>s$.

Eq. (\ref{intertw}) represents the intertwining relation  between
the free Hamiltonian $H_{0,0}$ and $H_{r,s}$. The eigenfunctions of
the latter are given by application
of the operator (\ref{intfree}) to the plane waves (\ref{plane}),
\begin{equation}\label{rline1}
    \psi^{\pm k}_{r,s}={\cal F}_{r,s} e^{\pm i k x}\, .
\end{equation}
The spectrum of $H_{r,s}$ is composed by a continuum of doubly
degenerate states with energies $E_k=k^2$, except for the $r+1$
states
\begin{equation}\label{rline2}
    \psi^{n}_{r,s}={\cal F}_{r,s}
    e^{i [n+u(r,s;n)]\, x}, \quad n=0,1,...,r \, ,
\end{equation}
which are singlets. The discrete
parameter $u(r,s;n)$ is defined here in the same way as in equation
(\ref{intfree}). The energy values of the singlet states
(\ref{rline2}) are
\begin{equation}\label{singener}
    E_{r,s;n}=
    \left(n +u(r,s;n)\right)^2 \,, \quad n=0,1,..,r\,.
\end{equation}
Clearly, the singlet states of the form (\ref{rline2}) correspond to
the states in (\ref{rline1}) taken for the particular values of $k$.
Two states obtained by the application of ${\cal F}_{r,s}$ to the
left and right moving plane waves with these special values of
$k\neq 0$ coincide modulo a
constant factor.  At the same time,
the application of the Crum-Darboux generator ${\cal F}_{r,s}$ to
the singlet state $\psi_0$ of the free particle produces a
nontrivial singlet ground state of the system $H_{r,s}$. We shall
return to this point below in the discussion of spectral
singularities in terms of the Wronskian of the solutions.

The picture can be understood alternatively by observing that  the
operator ${\cal F}_{r,s}$ annihilates $r$ states of the free
particle, which are complex linear combinations of the plane waves.
Namely,
\begin{eqnarray}
 {\cal F}_{r,s} \cos \gamma (x+i\delta)&=&0, \qquad
 \gamma=n+u(r,s;n)=\text{odd}\,, \label{anni1}\\
  {\cal F}_{r,s} \sin \gamma (x+i\delta)&=&0, \qquad
  \gamma=n+u(r,s;n)=\text{even}\,, \label{anni2}
\end{eqnarray}
with $n=1,2,...,r$ and $u(r,s;n)$ as defined above. Here the
resulting $r$ singlet states different from zero in the above
relations correspond to spectral singularities, they are located
inside the continuous spectrum. An additional singlet ground state
($E_0=0$) at the bottom of the spectrum has a different nature with
respect to the spectral singularities. States of this kind, being a
Crum-Darboux-transformed ground state of the free particle, also
appear in the Hermitian and non-Hermitian reflectionless potentials
\cite{susypt,susyscarf}. In the next section the difference between
the non-zero and zero energy singlet states of the Hamiltonian
(\ref{hrs}) will be clarified by applying a specific limit to
doubly-periodic finite-gap systems.

Performing a compactification in the coordinate as in the
 free particle case,
 the eigenfunctions are determined by the
 condition (\ref{condi}) imposed on the states (\ref{rline1}),
\begin{equation}\label{circle1}
    \psi^{\pm m}_{r,s}={\cal F}_{r,s} e^{\pm i m x}, \, \quad
    m=0,1,2,...
\end{equation}
 The solutions (\ref{circle1}) form an infinite
discrete set of wave eigenfunctions for doubly degenerate energy
levels, except the states (\ref{rline2}), which are singlets. Again,
we have $r+1$ singlet states, one of which is the ground state of
zero energy while the rest are exceptional points.

This example, provided by the compactified  ${\cal PT}$-symmetric
systems, reveals a subtlety related to the definition of spectral
singularities  and exceptional points in a periodic case. In such
class of the systems, they both have the same origin since the
spectral singularities transform into exceptional points just by
changing the topology of the quantum problem.

It is  useful to look at the peculiarity of these states from the
viewpoint of the Wronskian. For the physical wavefunctions
(\ref{rline1}) and (\ref{circle1}) we have
\begin{equation}
    W[\psi^{+ k}_{r,s},\psi^{- k}_{r,s}]=-2i
    k \prod_{n=1}^r(k^2-E_{r,s;n}) \, ,
\end{equation}
where for the case of the states (\ref{circle1}), the $k$ should be
replaced by $m$.   Being independent of $x$, the Wronskian vanishes
at the energies of the spectral singularities (\ref{singener}) as
well as at the lowest energy $E=0$ (which can be treated as a
trivial zero of $W$ corresponding to
the case of the coinciding arguments $\psi^{+
0}_{r,s}=\psi^{-0}_{r,s}$). This reflects linear dependence of the
pairs of states in (\ref{rline1}) and (\ref{circle1}). The second,
linear independent solutions of the stationary Schr\"odinger
equation for those energy values (including $E=0$) are not
periodic functions (being not
bounded in the non-compact topology case), and they do not belong
to the physical spectrum of $H_{r,s}$. In ref. \cite{samsonovroy},
Samsonov and Roy observed a similar phenomenon in the Wronskian for
the Hamiltonian $H_{1,0}(x+\zeta)$, where $\zeta$ is a complex
parameter. In their analysis, however, they imposed the boundary
conditions $\psi(\pi)=\psi(-\pi)=0$, which eliminate the existence
of spectral singularities in the spectrum, producing an infinite
number of discrete singlet states; this can be compared with the
Hermitian case $\delta=0$ we discussed at the beginning of the
section. This provides a further example of the importance of the
topology in Hermitian and non-Hermitian Hamiltonians.

\begin{figure}[h!]
\centering
\includegraphics[scale=0.8]{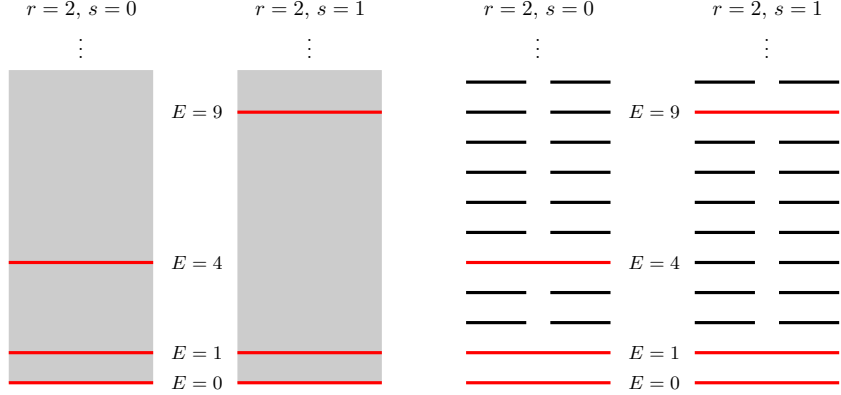}\caption{The
spectra for $r=2, s=0$ and $r=2,s=1$ for the Hamiltonians  with the
non-compact (left) and compact (right) coordinate. The spectrum in
the former case is a half-bounded  infinite continuum, without
spectral singularities above $E=4$ for $r=2, s=0$ and  $E=9$ for
$r=2,s=1$, while in the latter case there is an
infinite  number of degenerate
discrete states above those values corresponding to exceptional
points.} \label{fig2}
\end{figure}

The differences in the spectrum for
the discussed family of periodic systems  with compact and
non-compact topologies, and the comparison with the free particle
are illustrated by Fig. \ref{fig2} for the cases $r=2, s=0$ and
$r=2,s=1$.

In this section we explained the appearance  of spectral
singularities in the spectrum of $H_{r,s}$ by exploiting their
relation with the free particle system by means of Crum-Darboux
transformations. The non-degenerate nature of the spectral
singularities can be understood taking into account a peculiar
feature of the action of Crum-Darboux generators. The operator
(\ref{intfree}) annihilates $r$ states in the spectrum of $H_{0,0}$,
see the relations (\ref{anni1}) and (\ref{anni2}). Hence, one
naturally may expect the presence of singlet states, (\ref{rline2}),
in the spectrum of the intertwined Hamiltonian $H_{r,s}$, whose
energies coincide with the energy values of the annihilated states
of the free Hamiltonian. In other words, we say that the operators
(\ref{intfree}) remove every time one state located at the level
$E_{r,s;n}\neq 0$ from a doublet in the free particle spectrum,
generating a spectral singularity in the spectra of Hamiltonians
(\ref{hrs}) at the same energy
$E_{r,s;n}\neq 0$.

In next sections we will see that the origin of the singlet states,
in particular the spectral singularities, and the specific
degeneracy in the spectrum of $H_{r,s}$
have a remarkable interpretation
from the point of view of finite-gap potentials and a hidden
bosonized nonlinear supersymmetry.

\section{Darboux-Treibich-Verdier potentials and the origin of spectral singularities}
\label{sec3}

Generically, for a quantum periodic system free of singularities the
spectrum is composed by an infinite number of bands and gaps. By the
oscillation theorem \cite{osci}, the
number  of nodes of the band-edge states within the period interval
of the potential increases when energy increases. In the case of
analytical potentials the width of the gaps decreases exponentially
with increasing of the energy. There
exists, however, an important  class
of the systems for which
the number of bands and gaps is finite;
the corresponding potentials are known as finite-gap. One example of
a regular Hermitian finite-gap potential is provided by the family
of associated  Lam\'e potentials,
\begin{eqnarray}
    V^{AL}_{r,s}(x)
     &=& s(s+1)k^{2}\mathrm{sn}^{2}x+r(r+1)
    k^{2}\mathrm{sn}^{2}(x+K) \label{dtv}\\
    &=&
    s(s+1)k^{2}\mathrm{sn}^{2}x+
    r
    (r+1)%
        \frac{k^{ 2}
        \mathrm{cn}^{2}x}{\mathrm{dn}^{2}x}\,.\label{al2}
\end{eqnarray}
The potential is expressed in terms of the doubly periodic Jacobi
elliptic functions~\footnote{In (\ref{al2}), the properties of the
Jacobi elliptic functions under the real half-period displacement
were used, see \cite{WW}.} $\mathrm{sn}\, (x,k)$, $\mathrm{cn} \,
(x,k)$ and $\mathrm{dn} \, (x,k)$. Hereafter we will not display the
dependence on the modular parameter $0<k<1$ in
them.
The potential (\ref{dtv}) has a real,
$2K$, and an imaginary, $2iK'$, periods, where
$K=K(k)$ is the elliptic complete
integral of the first kind and $K'=K(k')$, $k'=\sqrt{1-k^2}$
\cite{WW}. The finite-gap nature of the potentials happens when both
parameters $r$ and $s$ take integer values. Specifically, if we take
$r>s$, the spectrum has exactly $r$  gaps (the case $r=s$ reduces to
the $s=0$ case with a real period $K$, see \cite{trisusyJpha}).
This property is correlated with the fact that the potentials
(\ref{dtv}) satisfy the non-linear stationary equation of order $r$
of the Korteweg de-Vries hierarchy.

When the modular parameter tends to the limit values, we obtain two
systems with essentially different spectra. In the limit $k
\rightarrow 0$, when the imaginary period turns into infinity, any
system with a finite number of gaps is reduced to the free particle.
The gaps between all the allowed bands disappear, and two  states of
the same periodicity and number of nodes at the edges of a gap
transform into two different states of the same energy; these
correspond to sine and cosine combinations of the plane-wave states
of the free particle. In the other limit $k \rightarrow 1$, when the
real period tends to infinity, the system is transformed into the
hyperbolic reflectionless
P\"oschl-Teller potential, with finite number of bound states equals
to the number of gaps. The valence bands shrink, and each pair of
band-edge states of the same valence band coalesces forming a unique
bound state.  Both described situations happen in a generic case of
Hermitian finite-gap potentials.  In the case of complex potentials,
the picture radically changes,  and leads to the origin of spectral
singularities.

A regular  complex finite-gap potential can be obtained by a complex
displacement of the coordinate in (\ref{dtv}). Performing the shift
in the half of the imaginary period plus for an imaginary constant
$i\delta$, $0<\delta< K'$, $x
\rightarrow x+iK' +i\delta$, the potential becomes
\begin{equation}\label{als}
    V^{ALS}_{r,s}
    (x)\equiv V^{AL}_{r,s}(x+iK'+i\delta)=r(r+1)
    \frac{\mathrm{dn}^{2}(x+i\delta)}
    {\mathrm{cn}^{2}(x+i\delta)}+
    {s(s+1)}\frac{1}
    {\mathrm{sn}^{2}(x+i\delta)%
    } \, .
\end{equation}
This potential is ${\cal PT}$-symmetric and the Hamiltonian
$H^{ALS}_{r,s}=-\frac{d^2} {dx^2}+V^{ALS}_{r,s}$  is a doubly
periodic generalization of (\ref{hrs}).  When $\delta=0$,  the sum
of (\ref{dtv}) and (\ref{als})
gives rise to the well-known family
of Darboux-Treibich-Verdier potentials \cite{Veselov},
\begin{equation}
    V^{DTV}=V^{AL}_{r',s'}(x)+V^{ALS}_{r,s}(x) \, .
\end{equation}

The limit cases of the modular parameter give us the systems with a
single, real or pure imaginary,
period,
\begin{eqnarray}
    H^{ALS}_{r,s} &\xrightarrow[k\rightarrow{}0]\,&-\frac{d^2}
    {dx^2}+\frac{r(r+1)}{\cos^2 (x+i\delta)}+\frac{s(s+1)}
    {\sin^2(x+i\delta)}\,, \\
    H^{ALS}_{r,s} &\xrightarrow[k\rightarrow{}1]\,&-\frac{d^2}
    {dx^2}+\frac{s(s+1)}
    {\tanh^2(x+i\delta)} +r(r+1) \, .
\end{eqnarray}

First we
discuss the spectral properties of complex finite-gap potentials
with the non-compact topology.

The limit $k \rightarrow 1$ produces, in analogy with the Hermitian
case, complex reflectionless Hamitonians
(with a single imaginary period)
that have bound states in their spectra \cite{susyscarf, bq}.

Surprisingly, another limit gives us something not noticed before in
the literature. When $k \rightarrow 0$, the gaps shrink
and disappear, and the band edge
states at the edges of the same energy gap coalesce into one singly
periodic state producing a spectral singularity.

As an example of this peculiar situation, let us consider the case
$r=2, s=0$, which corresponds to a $2$-gap system. The band-edge
states are given by
\begin{equation}
    \psi_0 = 1+k^2 +\sqrt{1-k^2+k^4}-\frac{3}{\mathrm{sn}^2\,(x+i\delta)}
    \,,
\end{equation}
\begin{equation}
    \psi_1 = \frac{\mathrm{cn}\,(x+i\delta)\mathrm{dn}\,(x+i\delta)}
    {\mathrm{sn}^2\,(x+i\delta)}\,, 
    \qquad \psi_2 = \frac{\mathrm{cn}\,(x+i\delta)}
    {\mathrm{sn}^2\,(x+i\delta)}\,, 
    \qquad \psi_3 = \frac{\mathrm{dn}\,(x+i\delta)}
    {\mathrm{sn}^2\,(x+i\delta)}\,,
\end{equation}
\begin{equation}
    \psi_4 = 1+k^2 -\sqrt{1-k^2+k^4}-
    \frac{3}{\mathrm{sn}^2\,(x+i\delta)}\,,
\end{equation}
and their energies are
\begin{equation}
    E_0 = 2\left( 1+k^2-\sqrt{1-k^2+k^4}\right)\,,
\end{equation}
\begin{equation}
    E_1 = 1+k\,,\qquad
    E_2 = 1+4 k\,,\qquad
    E_3 =4+k\,,
\end{equation}
\begin{equation}
    E_4 = 2\left( 1+k^2+\sqrt{1-k^2+k^4}\right)\,.
\end{equation}

In the limit $k=1$, the first valence band between $E_0$ and $E_1$
disappears, edge state $\psi_0=1/\sinh^2(x+i\delta)$ coincides up to
a multiplicative constant with $\psi_1$, and forms a bound state.
The same happens with the second valence band between $E_2$ and
$E_3$, and for $\psi_2=\psi_3=\cosh(x+i\delta)/\sinh^2(x+i\delta)$.
Quasi-periodic Bloch states of the conduction band are transformed
into scattering states of the continuous spectrum, at the bottom of
which is located the state  $\psi_4=1-3/\tanh^2(x+i\delta)$. The
described picture is typical for Hermitian finite-gap potentials in
the real infinite period limit.

Instead, in the limit when the complex period tends to infinity,
$k=0$, the state
$\psi_0=2-3/\sin^2(x+i\delta)$ is
located at the bottom of the continuum. The remaining pairs of band
edge states, $\psi_1=\psi_2=\cos(x+i\delta)/\sin^2(x+i\delta)$ and
$\psi_3=-3\psi_4=1/\sin^2(x+i\delta)$, coincide in the form of
spectral singularities.  In this limit, the quasi-periodic states
inside the valence and conduction bands transform  into periodic
states, whose period depends on energy. In the case with the compact
coordinate, the singlet states transform in the same way discussed
above, while the condition
(\ref{condi}) selects the periodic states of a fixed period $2\pi$
to be the double period of the
resulting potential.

\begin{figure}[h!]
\centering
\includegraphics[scale=0.8]{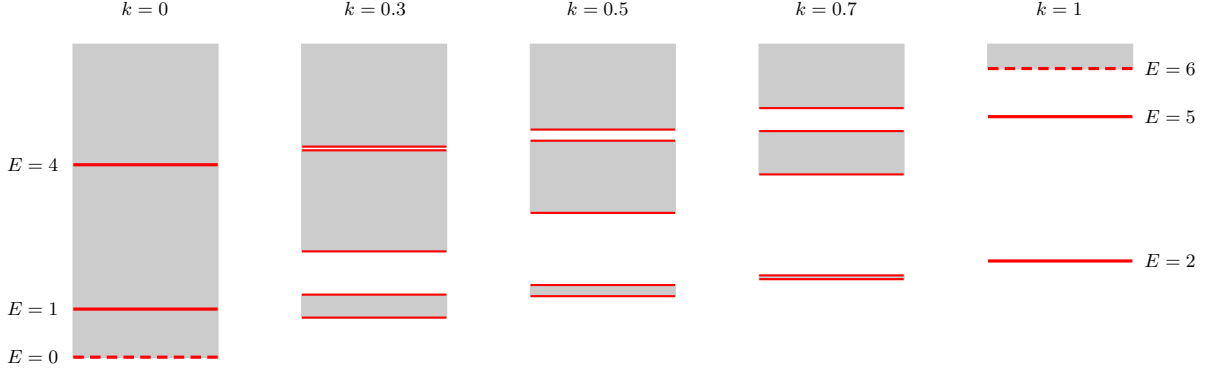}\caption{Spectra
of the 2-gap system $H_{2,0}^{ALS}$ are shown for the values of
the modular parameter $k=0, 0.3, 0.5, 0.7, 1$. In the limit when the
imaginary period tends to infinity, $k=0$, the system $H_{2,0}$ is
recovered. The spectral singularities (solid lines) appear at
energies $E=1$ and $E=4$ when two band-edge states with the same
periodicity, separated by a gap, coalesce, and the gap between
them vanishes. In the other limit $k=1$, a system with a pure
imaginary periodicity is obtained; two bound states (solid lines)
appear at energies $E=2$ and $E=5$. The singlet state at the
bottom of the continuous spectrum (dashed line) at energy $E=6$
($E=0$) corresponds to the limit $k=1$ ($k=0$); it has a nature of
the eigenfunction obtained by the application of the Crum-Darboux
operator (\ref{intfree}) to the free particle state $\psi=1$. The
system in the $k=1$ limit corresponds to the complexified Scarf II
potential \cite{susyscarf,bq}.} \label{fig3}
\end{figure}

 Fig. \ref{fig3} illustrates this
example, showing how the energies of the band-edge states are
transformed into corresponding singlets, particularly, into spectral
singularities when $k=0$. It is worth to note that periodic ${\cal
PT}$-symmetric potentials of the Darboux-Treibich-Verdier family
were treated before in ref.
\cite{KS}, but the existence of the spectral singularities in the
$k=0$ limit was not noticed.

\section{Hidden supersymmetry and spectral singularities}
\label{sec4}

The Hamiltonians $H_{r,s}$ display a
finite number of spectral singularities which appear as singlet
energy levels immersed into doubly degenerate continuous spectrum in
the non-compact topology case. The same happens with exceptional
points in the discrete spectrum of the compactified systems. These
features, the presence of several singlet states and double
degeneration of the rest of energy levels, are typical for a hidden,
bosonized (due to the absence of the spin degrees of freedom)
non-linear supersymmetry \cite{bosonized}. This kind of symmetry was
observed earlier  in the reflectionless systems with real
\cite{susypt} and complex \cite{susyscarf} potentials, as well as in
Hermitian periodic finite-gap \cite{trisusyprl,trisusyJpha} systems.
One can expect therefore that such a peculiar supersymmetry can also
be associated with the spectral singularities  and exceptional
points in the described class of the $\mathcal{PT}$-symmetric
systems.

Indeed, from the viewpoint of the limit of finite-gap potentials
(\ref{als}) discussed above, one knows that each potential
$V^{ALS}_{r,s}$ satisfies a corresponding  stationary non-linear
equation of the Korteweg-de Vries hierarchy. This fact implies the
existence of a nontrivial integral of motion ${\mathbb A}_{2r+1}$ of
differential order $2r+1$, which underlies the non-linear nature of
the hidden supersymmetry. Together with the Hamiltonian $H_{r,s}$,
the operator ${\mathbb A}_{2r+1}$ composes the Lax pair
\cite{Finitegap,Lax},
\begin{equation}\label{Adef}
    [{\mathbb A}_{2r+1},H_{r,s}]=0,
    \qquad -{\mathbb A}_{2r
    +1}^2=P(H_{r,s})  \, ,
\end{equation}
where $P(H_{r,s})$ is an order $2r+1$ (spectral) polynomial in the
Hamiltonian. Using the definition of the operator (\ref{dop}), one
can identify the operator ${\mathbb A}_{2r+1}$ as follows,
\begin{equation}\label{aop}
 {\mathbb A}_{2r+1}={\cal F}_{r,s}\frac{d}{dx}{\cal F}_{r,s}^T\,,
\end{equation}
that is nothing else  as a Crum-Darboux  dressed free particle
integral $\frac{d}{dx}$ \cite{GN}.  The transposition  $T$,
substituting here Hermitian conjugation in the case of a Hermitian
Hamiltonian, is defined by inversion of the order of first order
operator multipliers in Eq. (\ref{intfree}) accompanied by the
change $\frac{d}{dx}\rightarrow -\frac{d}{dx}$.

The remarkable property of the class
of the systems described by $H_{r,s}$ is related with the physical
sense of the operator (\ref{aop}), where the ${\cal PT}$-symmetry
plays a key role. In the Hermitian limit of the system when
$\delta=0$, the action of the operator ${\mathbb A}_{2r+1}$ on
physical states is ill-defined. As was noted in Section \ref{sec2},
when $\delta$ vanishes, the potential in $H_{r,s}$ becomes real,
having singularities in the real line. As a result, the appropriate
quantum treatment is such that the wave functions vanish at singular
points, and an infinite number of the discrete singlet bound states
appears. Though the higher order Lax operator is still commuting
with Hermitian Hamiltonian, its action on physical (bound) states
produces non-physical states,
\begin{equation}
    \delta=0 \quad \longrightarrow \quad {\mathbb
    A}_{2r+1}\psi_{\text{physical}}= \psi_{\text{non-physical}}
    \,.
\end{equation}
Such a situation takes place also in the conformal mechanics model
given by the inverse square potential $n(n+1)x^{-2}$ \cite{conf},
that corresponds to a rational limit (when both, real and imaginary
periods tend to infinity) of finite-gap Hermitian periodic systems.
However, for $\delta \neq 0$ the ${\cal PT}$-symmetry provides  the
``cure'' for the operator ${\mathbb A}_{2r+1}$: the states
(\ref{rline1}) and (\ref{circle1}) are eigenstates of this operator,
\begin{equation}
    {\mathbb A}_{2r+1} \psi^{\pm k}_{r,s} =
   \pm i k \prod_{n=1}^r (k^2-
    E_{r,s;n}) \psi^{\pm k}_{r,s}, \quad  {\mathbb A}_{2r+1} \psi^{\pm m}_{r,s} =
  \pm  i m \prod_{n=1}^r  (m^2-
    E_{r,s,n}) \psi^{\pm m}_{r,s}  \, .
\end{equation}
Particularly, the non-linear
operator (\ref{aop}) annihilates all the singlet states. These are
spectral singularities (exceptional points) and the state in the
bottom of the spectrum in the systems with the non-compact
(compact) topology. This can be seen also taking the square of
${\mathbb A}_{2r+1}$,
\begin{equation}
    -{\mathbb A}_{2r+1}^2=H_{r,s}\prod_{n=1}^r (H_{r,s}-
    E_{r,s;n})^2 \, ,
\end{equation}
where the roots of the operator-valued polynomial are
the energies of the singlet states. In this sense we can say that
the ${\cal PT}$- symmetry restores the physical meaning of ${\mathbb
A}_{2r+1}$, which in the Hermitian case, $\delta=0$, was broken
\cite{pecu}.

The supersymmetric structure can be revealed by identifying the
supercharges as follows,
\begin{equation}
    Q_1=i{\mathbb A}_{2r+1}, \qquad Q_2=i\Gamma Q_1\,,
\end{equation}
where $\Gamma$ is a $\Z_2$-grading operator,
\begin{equation}
    \Gamma={\cal P} e^{-2i\delta\frac{d}{dx}},
    \qquad [\Gamma,H_{r,s}]=0,
    \qquad \{\Gamma,Q_a\}=0,
    \qquad
    \Gamma^2=1\,.
\end{equation}
Integral $\Gamma$, being $\mathcal{PT}$-symmetric, $[\mathcal{PT},
\Gamma]=0$, produces a pure imaginary shifting of the coordinate for
$-2i\delta$ followed by the action of the parity operator.  As the
grading operator  commutes with the $\mathcal{PT}$ operator, from
the definition of (\ref{aop}) and (\ref{intfree}) it
follows that the supercharges are
$\mathcal{PT}$-symmetric operators,
\begin{equation}
[\mathcal{PT}, Q_a]=0, \qquad a=1,2 \, .
\end{equation}
Acting on
the Hamiltonian eigenstates $\psi^{+ k}_{r,s}$ ($\psi^{+ m}_{r,s}$)
and $\psi^{- k}_{r,s}$ ($\psi^{- m}_{r,s}$) given by Eqs.
(\ref{rline1}) and (\ref{circle1}), operator $\Gamma$ transforms
them  mutually,
\begin{equation}
    \Gamma \psi^{\pm k}_{r,s}=(-1)^r e^{\pm 2 \delta k}
    \psi^{\mp k}_{r,s}, \qquad \Gamma
    \psi^{\pm m}_{r,s}=(-1)^r e^{\pm 2 \delta m}
    \psi^{\mp m}_{r,s} \, .
\end{equation}

 On the other hand, all
the singlet states are eigenstates of $\Gamma$. The corresponding
$N=2$ nonlinear superalgebra generated by the Hamiltonian $H_{r,s}$
and supercharges $Q_a$ reads as
\begin{equation}\label{susyalg}
    [Q_a,H_{r,s}]=0, \qquad \{Q_a,Q_b\}=2\delta_{ab}H_{r,s}\prod_{n=1}^r
    (H_{r,s}-
    E_{r,s;n})^2\,.
\end{equation}

 As follows from
(\ref{susyalg}), the nonlinear superalgebra  detects all the singlet
states in the spectrum; moreover, it distinguishes spectral
singularities from the  singlet ground state: unlike the energy
level $E=0$,  all the spectral singularities appear as double roots
of the polynomial in Hamiltonian.
The same holds for exceptional
points in the case of the compact topology.

\section{Discussion}
\label{sec5}

In this article we show from different points of view how spectral
singularities appear in ${\cal PT}$-symmetric singly periodic
finite-gap systems with non-compact topology.
These states correspond to
exceptional points when topology is changed for a compact one.

The examples discussed here test the effectiveness of Crum-Darboux
transformations for non-Hermitian Hamiltonians. Applying them to the
free particle Hamiltonian,  we construct the systems which display
singlet states inside the continuum. These states are known as
spectral singularities; these are a specific feature of complex
Hamiltonians.  We note here that the models defined by (\ref{hrs})
can be extended to a more  generic family of singly periodic
finite-gap systems by using complex Crum-Darboux transformations
different from those in
(\ref{intfree}).

 In ref.
\cite{susyscarf} it was shown that starting from the Hamiltonian
$H_{0,0}$, it is possible to obtain complex reflectionless
potentials by choosing non-physical states of the free system as a
kernel of the Crum-Darboux operator. In a similar way, selecting
physical solutions of the free
Hamiltonian, displaced for a complex
constant, as zero modes of the non-Hermitian Crum-Darboux
operators, complex systems with spectral singularities can be
constructed. Outside the scope of the present paper, an interesting
approach for the comprehension of this kind of states would be that
related to quasi-exact solvability, see refs. \cite{trisusyJpha,
trisusyprl, bqr}.

Non-Hermitian finite-gap potentials were analyzed from the point of
view of the corresponding infinite period complex and real limits.
We explain how the band-edge states coalesce and produce the
spectral singularities when the complex period is infinite. In this
picture the presence of spectral singularities is understood: it
corresponds to a remarkable, peculiar  feature of complex periodic
potentials, with no analog  in the Hermitian case. A more detailed
investigation on the hidden supersymmetries and related properties
of complex doubly periodic
finite-gap systems deserves a separate, further investigation
\cite{CJP}.

The existence of a finite number of spectral singularities leads to
an additional feature of the systems discussed in this paper. Their
non-degeneracy alongside with the
doubly degenerate continuum are
naturally explained by a hidden bosonized
non-linear supersymmetry, whose
structure also distinguishes spectral singularities from the singlet
ground state. The hidden non-linear supersymmetry, which is related
with the Lax pair of the KdV hierarchy, in the Hermitian limit, when
$\delta=0$, has a completely different nature. In such a limit the
integral of motion is ill defined, producing non-physical states. In
the sense we show that the pathologies of the hidden supersymmetry
in the Hermitian case, originated from real singularities, can be
circumvented by changing the Hermiticity property of the Hamiltonian
for the ${\cal PT}$-symmetry.

It would be interesting to study
the breaking of ${\cal PT}$-symmetry and the disappearance of
spectral singularities by appropriate modification of the systems
(\ref{hrs}). In the same
direction, the meaning of the investigated spectral singularities in
the context of optics and matter waves could be also a relevant
problem to investigate.

As a final remark, we note that it
is interesting to apply the ideas of the present paper to study the
models described by the first order Dirac-type Hamiltonians \cite{bdg},
particularly, to those related to the topologically nontrivial
solutions in the Gross-Neveu model \cite{BD,GN}, and to the physics
of nanotubes \cite{nanotubes}.

\vskip0.2cm

\noindent \textbf{Acknowledgements}

 The work has been partially
 supported by FONDECYT Grants 1095027 (MP)  and 3100123 (FC).
FC acknowledges also financial support via the CONICYT grants
79112034 and Anillo ACT-91: \textquotedblleft Southern Theoretical
Physics Laboratory\textquotedblright\ (STPLab).  MP and FC are
grateful, respectively, to CECs and Universidad de Santiago de Chile
for hospitality. The Centro de Estudios Cient\'{\i}ficos (CECs) is
funded by the Chilean Government through the Centers of Excellence
Base Financing Program of Conicyt.


\begin{thebibliography}{99}

\bibitem{B&B}
 C.~M.~Bender, S.~Boettcher,
Phys.\ Rev.\ Lett.\  {\bf 80}, 5243 (1998), [arXiv:physics/9712001].

 \bibitem{reviews}
For reviews, see:  C.~M.~Bender,
  Rept.\ Prog.\ Phys.\  {\bf 70}, 947  (2007), [arXiv:hep-th/0703096];

  A.~Mostafazadeh,
  Int.\ J.\ Geom.\ Meth.\ Mod.\ Phys.\  {\bf 7}, 1191 (2010), [arXiv:0810.5643 [quant-ph]].


 \bibitem{optics}
 A. Guo, G. J. Salamo, D.  Duchesne, R. Morandotti,  M. Volatier-Ravat,
 V. Aimez,  G. A. Siviloglou,  D. N. Christodoulides,
 Phys.\ Rev.\ Lett.\  {\bf 103}, 093902 (2009);
C.E. Ruter, K.G. Makris, R. El-Ganainy, D.N. Christodoulides, M.Segev, D. Kip,
 Nature Physics {\bf 6}, 192 (2010).

\bibitem{QFT}
  C.~M.~Bender, K.~A.~Milton and V.~M.~Savage,
  Phys.\ Rev.\ D {\bf 62}, 085001 (2000),
  [hep-th/9907045];
   C.~W.~Bernard and V.~M.~Savage,
  Phys.\ Rev.\ D {\bf 64}, 085010 (2001),
  [hep-lat/0106009];
  C.~M.~Bender, D.~C.~Brody and H.~F.~Jones,
  Phys.\ Rev.\ Lett.\  {\bf 93},  251601 (2004),
  [hep-th/0402011];
  Phys.\ Rev.\ D {\bf 70}, 025001 (2004)
  [Erratum-ibid.\ D {\bf 71}, 049901 (2005)],
  [hep-th/0402183];


\bibitem{grav}
  P.~D.~Mannheim,
  Found.\ Phys.\  {\bf 42}, 388 (2012),
  [arXiv:1101.2186 [hep-th]];
  \emph{``Astrophysical Evidence for the Non-Hermitian but $PT$-symmetric Hamiltonian of Conformal Gravity,''}
  [arXiv:1205.5717 [hep-th]]

\bibitem{rqm}
    J.~Smejkal, V.~Jakubsky and M.~Znojil,
  J.\ Phys.\ Stud.\  {\bf 11}, 45 (2007),
  [hep-th/0611287].
 C.~M.~Bender and P.~D.~Mannheim,
  Phys.\ Rev.\ D {\bf 84}, 105038 (2011),
  [arXiv:1107.0501 [hep-th]].






 \bibitem{kato}
T. Kato, Perturbation Theory for Linear Operators (Springer, New York, 1966).

 \bibitem{Naimark}
M. A. Naimark, Amer. Math. Soc. Transl. {\bf 16}, 103 (1960).

 \bibitem{EP1}
M.V. Berry, J. Mod. Opt. {\bf 50}, 63 (2003); Czech. J. Phys. {\bf 54}, 1039 (2004).

\bibitem{EP2}
W.D. Heiss, J. Phys. A {\bf 37}, 2455 (2004).

\bibitem{EP3}
C.~Dembowski, H.-D.~Gr\"af, H.~L.~Harney, A.~Heine, W.~D.~Heiss, H.~Rehfeld and A.~Richter,
  Phys.\ Rev.\ Lett.\  {\bf 86}, 787 (2001);
  W.~D.~Heiss, M.~Muller and I.~Rotter,
  Phys.\ Rev.\ E {\bf 58}, 2894 (1998);
  [arXiv:quant-ph/9805038];
C. Dembowski, B. Dietz, H.-D. Gr\"af, H. L. Harney, A. Heine, W. D.
Heiss   and A. Richter, Phys. Rev. E {\bf 69}, 056216 (2004); T.
Stehmann, W. D. Heiss and F. G. Scholtz,   J. Phys. A: Math. Gen.
{\bf 37}, 7813 (2004).

\bibitem{EP4}

A. A. Mailybaev, O. N. Kirillov and A. P. Seyranian, Phys. Rev. A
{\bf 72}, 014104 (2005).

\bibitem{EP5}

  B.~Dietz, H.~L.~Harney, O.~N.~Kirillov, M.~Miski-Oglu, A.~Richter and F.~Schafer,
  Phys.\ Rev.\ Lett.\  {\bf 106}, 150403 (2011),
  [arXiv:1008.2623 [nlin.CD]].

\bibitem{EP6}

H. Ramezani, T. Kottos, V. Kovanis and D. Christodoulides, Phys.
Rev. A {\bf 85}, 013818 (2012).

\bibitem{bqr}
  B.~Bagchi, C.~Quesne and R.~Roychoudhury,
  J.\ Phys.\ A A {\bf 41}, 022001 (2008),
  [arXiv:0710.1802 [quant-ph]].

\bibitem{Samsonov1}
B. F. Samsonov,  J. Phys. A: Math. Gen. \textbf{38},  L397 (2005).

\bibitem{AMHMD}
A.~Mostafazadeh and H.~Mehri-Dehnavi,
  J.\ Phys.\ A  {\bf 42}, 125303 (2009),
  [arXiv:0901.3563 [math-ph]].

  \bibitem{Longhi1}
  S. Longhi , 
  Phys. Rev. A {\bf 81}, 022102 (2010), [arXiv:1001.0962
  [quant-ph]].

\bibitem{AM1}
  A.~Mostafazadeh,
  Phys.\ Rev.\ Lett.\  {\bf 102}, 220402 (2009),
  [arXiv:0901.4472 [math-ph]].

\bibitem{Samsonov2}
  B.~F.~Samsonov,
  J.\ Phys.\ A  {\bf 38}, L571 (2005).

  \bibitem{Samsonov3}
  B.~F.~Samsonov,
  J.\ Phys.\ A  {\bf 43}, 402006 (2010),
  [arXiv:1006.0282 [math-ph]].

\bibitem{Samsonov4}
  B.~F.~Samsonov,
  J.\ Phys.\ A  {\bf 44}, 392001 (2011),
  [arXiv:1007.4421 [quant-ph]].

\bibitem{Samsonov5}
B.~F.~Samsonov,
\emph{``Hermitian Hamiltonian equivalent to a given non-Hermitian one. Manifestation of spectral singularity
''}, [arXiv:1207.2525 [math-ph]].

\bibitem{Andrianov:2010an}
  A.~A.~Andrianov, F.~Cannata and A.~V.~Sokolov,
  J.\ Math.\ Phys.\  {\bf 51}, 052104 (2010),
  [arXiv:1002.0742 [math-ph]].

  \bibitem{AM2}
  A.~Mostafazadeh,
  Phys.\ Rev.\ A {\bf 80}, 032711 (2009),
  [arXiv:0908.1713 [quant-ph]];
  Phys.\ Rev.\ A {\bf 83}, 045801 (2011),
  [arXiv:1102.4695 [physics.optics]];
 \emph{``Self-dual Spectral Singularities and Coherent Perfect Absorbing Lasers without PT-symmetry''},
  [arXiv:1205.4560 [quant-ph]].

  \bibitem{AM3}
  A.~Mostafazadeh and S.~Rostamzadeh,
  \emph{``Perturbative Analysis of Spectral Singularities and Their Optical Realizations''},
  [arXiv:1204.2701 [quant-ph]].


    \bibitem{AM4}
  A.~Mostafazadeh and M.~Sarisaman,
  \emph{``Optical Spectral Singularities and Coherent Perfect Absorption in a Two-Layer Spherical Medium''},
 [arXiv:1205.5472 [physics.optics]].

   \bibitem{AM5}
    A.~Mostafazadeh ,
 \emph{``Spectral Singularities Do Not Correspond to Bound States in the Continuum''}
 , [arXiv:1207.2278 [math-ph]].

  \bibitem{Heiss:2011ad}
  W.~D.~Heiss and R.~G.~Nazmitdinov,
  Eur.\ Phys.\ J.\ D {\bf 63}, 369 (2011),
  [arXiv:1105.3141 [cond-mat.quant-gas]].

\bibitem{Finitegap}
 E.~D.~Belokolos  et al, \emph{Algebro-Geometric Approach
to Nonlinear Integrable Equations}, (Springer, Berlin, 1994).

\bibitem{bosonized}
  M.~S.~Plyushchay,
  Annals Phys.\  {\bf 245}, 339 (1996),
  [arXiv:hep-th/9601116];
  Int.\ J.\ Mod.\ Phys.\ A {\bf 15}, 3679 (2000),
  [arXiv:hep-th/9903130].


\bibitem{MatSal}
V. B. Matveev and M. A. Salle, \emph{Darboux Transformations and
Solitons},  (Springer, Berlin, 1991).


 \bibitem{susypt1}
  F.~Cannata, G.~Junker and J.~Trost,
  Phys.\ Lett.\ A {\bf 246}, 219 (1998), [arXiv:quant-ph/9805085].

\bibitem{BDM}
  C.~M.~Bender, G.~V.~Dunne and P.~N.~Meisinger,
  Phys.\ Lett.\ A {\bf 252}, 272 (1999),
  [arXiv:cond-mat/9810369].

\bibitem{jones}
H.F Jones, Phys. Lett. A {\bf 262}, 242 (1999).

\bibitem{cervero}
J. M. Cerver\'o,
Phys. Lett. A {\bf 317}, 26 (2003) .

\bibitem{shin}
K.C. Shin,
J. Phys. A: Math. Gen. {\bf 37}, 8287 (2004), [arXiv:math-ph/0404015].

\bibitem{samsonovroy}
B. F. Samsonov and  P. Roy,
  J. Phys. A: Math. Gen. {\bf 38}, L249 (2005), [arXiv:quant-ph/0503040].


\bibitem{Sirre}
B.F. Samsonov, Phys. Lett. A {\bf 358}, 105 (2006), [arXiv:quant-ph/0602101].

\bibitem{kharemandal}
A. Khare and B. P. Mandal, Pramana J. Phys. {\bf 73}, 387 (2009),
[arXiv:1112.3767 [math-ph]].



\bibitem{Longhi2}
 S. Longhi ,
    J. Phys. A: Math. Theor. {\bf 44}, 485302 (2011), [arXiv:1111.3448 [quant-ph]].

  \bibitem{self}
  G.~V.~Dunne and J.~Feinberg,
  Phys.\ Rev.\ D {\bf 57}, 1271 (1998), [arXiv:hep-th/9706012].


 \bibitem{susypt}
  F.~Correa and M.~S.~Plyushchay,
  Annals Phys.\  {\bf 322}, 2493 (2007), [arXiv:hep-th/0605104].

  \bibitem{susyscarf}
  F.~Correa and M.~S.~Plyushchay,
  Annals Phys.\  {\bf 327}, 1761 (2012),
  [arXiv:1201.2750 [hep-th]].

\bibitem{osci}
W. Magnus and S. Winkler, \emph{Hill's equation}, (Wiley, New York, 1966).

\bibitem{WW}
E. T. Whittaker and G. N. Watson, \emph{Course of Modern
Analysis}, (Cambridge Univ. Press, Cambridge, 1980).

\bibitem{trisusyJpha}
F.~Correa, V.~Jakubsky and M.~S.~Plyushchay,
  J.\ Phys.\  A {\bf 41}, 485303 (2008), [arXiv:0806.1614 [hep-th]].

    \bibitem{Veselov}
A. P. Veselov, 
Lett. Math. Phys. {\bf 96}, 
209 (2011), [arXiv:1004.5355].

\bibitem{bq}
  B. Bagchi and C. Quesne,
  Phys. Lett. A {\bf 273}, 285  (2000), [arXiv:math-ph/0008020];
J.\ Phys.\ A {\bf 43}, 305301 (2010), [arXiv:1002.4309 [quant-ph]].

\bibitem{KS}
A. Khare and Uday Sukhatme, Phys. Lett. A {324}, 406 (2004),
 [arXiv:quant-ph/0402106]; J. Math. Phys. {\bf 46}, 082106 (2005), [arXiv:math-ph/0505027];
  J. Math. Phys. {\bf 47}, 062103 (2006), [arXiv:quant-ph/0602105].

  \bibitem{trisusyprl}
  F.~Correa, V.~Jakubsky, L.~-M.~Nieto and M.~S.~Plyushchay,
  Phys.\ Rev.\ Lett.\  {\bf 101}, 030403 (2008), [arXiv:0801.1671 [hep-th]].

\bibitem{Lax}
  P.~D.~Lax,
  Commun.\ Pure Appl.\ Math.\  {\bf 21}, 467 (1968).

\bibitem{GN}
M.~S.~Plyushchay and L.~-M.~Nieto,
  Phys.\ Rev.\ D {\bf 82}, 065022 (2010),
  [arXiv:1007.1962 [hep-th]];
  M.~S.~Plyushchay, A.~Arancibia and L.~-M.~Nieto,
  Phys.\ Rev.\ D {\bf 83}, 065025 (2011),
  [arXiv:1012.4529 [hep-th]].

 \bibitem{conf}
F.~Correa, M.~A.~del Olmo and M.~S.~Plyushchay,
  Phys.\ Lett.\ B {\bf 628}, 157 (2005),
  [arXiv:hep-th/0508223].


\bibitem{pecu}
F.~Correa and M.~S.~Plyushchay,
  J.\ Phys.\ A {\bf 40}, 14403 (2007),
  [arXiv:0706.1114 [hep-th]].

\bibitem{CJP}
F.~Correa, V.~Jakubsky and M.~S.~Plyushchay, \emph{in preparation.}

\bibitem{bdg}
 F.~Correa, G.~V.~Dunne and M.~S.~Plyushchay,
  Annals Phys.\  {\bf 324}, 2522 (2009),
  [arXiv:0904.2768 [hep-th]].

\bibitem{BD}
G.~Basar and G.~V.~Dunne,
  Phys.\ Rev.\ Lett.\  {\bf 100}, 200404 (2008),
  [arXiv:0803.1501 [hep-th]];
  Phys.\ Rev.\ D {\bf 78}, 065022 (2008),
  [arXiv:0806.2659 [hep-th]].


\bibitem{nanotubes}
V.~Jakubsky and M.~S.~Plyushchay,
  Phys.\ Rev.\ D {\bf 85}, 045035 (2012),
  [arXiv:1111.3776 [hep-th]].



\end{thebibliography}
\end{document}